\documentclass[11pt]{article}
\usepackage[utf8]{inputenc}
\usepackage[a4paper,left=3cm,right=3cm,top=3.0cm,,bottom=3.0cm]{geometry}
\usepackage{cite}
\usepackage{amsmath,amssymb,amsfonts}
\usepackage{graphicx}
\usepackage{tikz}
\usetikzlibrary{positioning}
\usepackage[dvipsnames]{xcolor}
\usepackage{circuitikz}
\usepackage{caption}
\usepackage{booktabs}
\usepackage{subcaption}
\usepackage{textcomp}
\usepackage{xcolor}
\usepackage{booktabs}
\usepackage{pgfplots}
\usepackage{bbold}
\usepackage{siunitx}
\usepackage{subcaption}
\usepackage{algorithm}
\usepackage{algpseudocode}
\usepackage{enumitem}
\usepackage{multirow}
\usepackage{multicol}
\usepackage{hyperref}
\usepackage{authblk}

\def\BibTeX{{\rm B\kern-.05em{\sc i\kern-.025em b}\kern-.08em
    T\kern-.1667em\lower.7ex\hbox{E}\kern-.125emX}}

\title{Identifiability of Low Frequency Li-ion Battery Parameters in Time Domain\\[14cm]}
\author{Vladimir Sovljanski\textsuperscript{a,*}, Mario Paolone\textsuperscript{a}\\
\href{mailto:vladimir.sovljanski@epfl.ch}{vladimir.sovljanski@epfl.ch}, \href{mailto:mario.paolone@epfl.ch}{mario.paolone@epfl.ch}}
\affil{\textsuperscript{a}Distributed Electrical Systems Laboratory\\ \'Ecole Polytechnique F\'ed\'erale de Lausanne\\
Lausanne, Switzerland}

\date{}
\begin{document}

\begin{titlepage}
\maketitle
\noindent

\thispagestyle{empty}

\vspace{-45pt}
\begin{flushleft}
\large
\textsuperscript{*}Corresponding author\\
EPFL STI IEM DESL\\
Bâtiment ELL\\
Station 11\\
CH-1015 Lausanne
\end{flushleft}

\centering{
\date{\today}}
\end{titlepage}

\begin{abstract}
    
This paper investigates the identification of observable low-frequency (LF) parameters of battery cell's equivalent circuit models (ECMs) using time-domain voltage and current measurements sampled at low frequency by built-in battery management systems (BMS) during operation. Accurate estimation of such parameters is challenging due to measurement resolution available in practical settings.

To address this, a modeling and identification framework is proposed in which fractional constant phase element (CPE), commonly used to model LF diffusion phenomena of battery cells,  is approximated in the time domain using a high-order RC network with a recursive definition. The parameter estimation problem is formulated as a constrained, non-convex least-squares problem in a discretized state-space representation. To improve robustness, parameter initialization strategies, bounds, and a procedure for selecting the number of RC branches are rigorously derived.

The method is evaluated in a numerical study based on a power system application where the battery under the study provides primary frequency control to the grid. Under noise levels representative of typical BMS measurements, the proposed approach achieves, from time-domain measurements, accurate LF parameter estimation (including the CPE), with average errors below 1\%.
\end{abstract}

\pagebreak

\section{Introduction}\label{sec:Intro}

\setcounter{page}{1}

During operation, the built-in battery management system (BMS) present in any battery system typically monitors pack and module voltage, current and temperature, estimates the state of charge (SOC) and state of health (SOH). In addition, the BMS performs several critical functions, including cell balancing and various protection mechanisms. However, obtaining more detailed parameters for battery equivalent circuit models (ECMs) generally requires performing additional measurements, such as electrochemical impedance spectroscopy (EIS) \cite{shangshang_electrochemical_2021}. Since current commercial BMSs lack the instrumentation needed to perform such tests, EIS measurements typically require disassembling the battery pack or modules and connecting them to specialized equipment. As a result, EIS is usually conducted under laboratory conditions when the battery is not in operation, although onboard EIS is emerging as a promising feature for future advanced BMS technologies \cite{luo_review_2026}.

Commercial built-in BMS in EVs often operate with low sampling frequencies for monitoring cell voltages and currents, which can be as low as 1 Hz \cite{kulkarni_advanced_2025}. In \cite{berger_benchmarking_2024}, sampling rates of 100 Hz for current and 10 Hz for voltage are reported as typical requirements for BMS algorithms. Measurements of voltage, current, and temperature may have higher sampling frequencies in case they are used for safety/protection purposes. Although strict requirements are not explicitly defined, \cite{tadoum_standards_2025} suggests that establishing minimum sampling frequency standards would improve the performance of BMS algorithms as demonstrated experimentally in \cite{gu_influence_2019}. Low sampling frequencies clearly impact the observability of ECM parameters \cite{zhou_novel_2020}. 
            
In this paper, we consider the estimation of observable ECM parameters using LF sampled data from the BMS. The authors of \cite{iurilli_use_2021} summarize ECM representations across different frequency regions reported in the literature. As noted, the LF behavior, when included in the model, is typically represented by the Warburg impedance or, more generally, by a constant phase element (CPE). As a fractional-order circuit component, the CPE is well suited for frequency-domain analysis. However, in the time domain, it leads to circuit equations involving derivatives and integrals of non-integer order.
            
As known, an exact ECM representation of CPE in time domain would require an infinite number of RC branches. Nevertheless, over a limited frequency range, it can be accurately approximated using a finite number of RC elements \cite{plett_battery_2015}. The works of Valsa \cite{valsa_network_2011}, \cite{valsa_rc_2013} introduce a method for approximating a general CPE using a network of RC branches with a recursive definition, as in \cite{tenreiro_machado_discrete-time_2001}, complemented by additional corrective resistive and capacitive elements to achieve an accurate approximation over a specified frequency range of interest. This approach is directly applied in \cite{fonseca_state-space_2020} to derive state-space representations of battery models, which are widely used in the literature (e.g., \cite{sossan_achieving_2016}, \cite{alavi_identifiability_2017}).
            
In \cite{santos_modeling_2023}, the authors employ a seventh-order RC circuit and derive a finite-order linear time-invariant (LTI) state-space approximation of the Warburg impedance using the Ho–Kalman algorithm. The parameters of the approximated circuit are estimated from time-domain voltage and current measurements, using pulse excitation. In \cite{alavi_time-domain_2015}, the authors demonstrate parameter estimation for an ECM consisting of a resistor in series with a Zarc element (i.e., a resistor in parallel with a CPE), using pseudo-random binary sequence (PRBS) and multisine excitation signals spanning a wide frequency range. Their approach relies on fractional-order system identification, presented in \cite{cois_fractional_2001}.
            
Similarly, \cite{adel_time-domain_2025} proposes a two-stage method for estimating the parameters of a battery ECM composed of a resistor, Zarc element, and CPE, based on fractional-order system identification with non-zero initial conditions under PRBS excitation across a wide frequency range. Ref. \cite{schroer_adaptive_2020} investigates parameterizations across different frequency regions using both frequency-domain (EIS) and time-domain (pulse profile) methods. Finally, in \cite{xia_accurate_2016}, the authors estimate the parameters of a second-order RC circuit using continuous-time system identification and an instrumental variable method, with pulse excitations of varying amplitudes.

In view of the above state of the art on ECM parameter identification, most works focusing on time-domain parameter estimation consider models with a limited number of RC branches to represent the battery. Moreover, they rely on pre-engineered and identification friendly excitation signals such as pulses, PRBS, or multisine signals, which are typically artificially imposed rather than obtained from the battery’s normal operating conditions. The use of such signals requires the interruption of the BESS operation just to carry the identification procedure making their adoption not adequate for a continuous monitoring of the battery ECM.
    
In this work, we investigate the identification of observable (LF) ECM parameters using time-series measurements of voltage and current sampled at low frequency by a built-in BMS, during normal battery operation. To model the ECM components observable from LF-sampled BMS data in the time domain, we employ an ECM structure consisting of a resistance and a high number of RC branches with a recursive definition to approximate a CPE (or equivalently, the Warburg impedance).

The parameter estimation problem is formulated as a constrained, non-convex least-squares (LS) minimization problem, expressed in a standard discretized state-space form. To improve computational efficiency and mitigate issues related to non-convexity, we propose suitable parameter initialization strategies and bounds that effectively reduce the solution space. Furthermore, we design an algorithmic approach to determine a sufficient number of RC branches, thereby enhancing convergence properties.

The estimated parameters of the resulting ECM enable the recovery of the corresponding CPE parameters without requiring explicit fractional-order system identification.

This paper is organized as follows: in Section \ref{sec:method}, we present the method including problem formulation using state-space representation (with approximated CPE), parameter initialization, estimation and reconstruction of original CPE parameters. Section \ref{sec:results} presents the results of the numerical study including a power system use case of battery providing primary frequency control. Section \ref{sec:Conclusion} concludes the paper.

\section*{Keywords}
Li-ion batteries, BMS, online parameter estimation, Constant-phase element, Warburg impedance, least squares.
\newpage
\section{Method}\label{sec:method}

\subsection{Battery Equivalent Circuit at Low-Frequencies}\label{sec:BatteryECM_at_LF}

The ECM of a battery cell commonly used in literature is shown in Fig 1. As known, it consists of elements representing high-, mid- and low-frequency behavior. The high-frequency (HF) part is composed of series resistance $R_s$ connected in series to HF constant phase element (CPE) with parameters $Q_{HF}$ and $\phi_{HF} \in \left[-1,0\right)$ (note that for $\phi_{HF} = -1$ the CPE becomes a pure inductor with $L=Q_{HF}^{-1}$). The mid-frequency (MF) part is modeled with one or multiple parallel branches of resistance and CPE. The LF part is usually modeled as a CPE or the Warburg impedance, (which is a special case of LF CPE the exponent $\phi=0.5$)\footnote{Using the CPE with $\phi$ as a non-fixed parameter can improve the LF fitting, typically done from EIS measurements in frequency domain}. 

In general, not all the parameters from the complete ECM are observable from measured battery voltage and current. The reasons for this can be both limited sampling frequency of reported battery voltage and current and type of the excitation signal. In this paper we assume that: (1) BMS has a low sampling frequency (in order of seconds); (2) measurements are collected from the BMS while the battery is under operation (i.e., we are not allowed to excite the system with specific signals such as PRBS or multisine) in order to continuously infer ECM parameters. Therefore, in what follows we analyze the ECM of observable parameters at LF.

\begin{figure}[h]\label{fig:Full_ECM}
    \centering
    \begin{subfigure}{0.8\textwidth}
        \centering
        \begin{center}\begin{circuitikz}\draw
          (0,3) to[R, l=\mbox{$R_{\text{HF}}$}] ++(2.5,0)
          (2.5,3) node[circ]{}
          (0,3) node[circ]{}
          (2.5,3) to[european resistor, l=\mbox{$Q_{\text{HF}},\phi_{\text{HF}}$}, a={$\bar{Z}_{\text{CPE}}^{\text{HF}}$}] ++(2.5,0)
          (5,3) node[circ]{}
          (5,3) to[short] ++(0,1)
          to[short] ++(0,-2)
          (5,4) to[resistor, l=\mbox{$R_{\text{MF}}$}] ++(2.5,0)
          (5,2) to[european resistor, l=\mbox{$Q_{\text{MF}},\phi_{\text{MF}}$},
          a={$\bar{Z}_{\text{CPE}}^{\text{MF}}$}] ++(2.5,0)
          (7.5,4) to[short] ++(0,-2)
          (7.5,3) node[circ]{}
          (7.5,3) to[european resistor,
          l=\mbox{$Q_{\text{LF}},\phi_{\text{LF}}$},
          a={$\bar{Z}_{\text{CPE}}^{\text{LF}}$}] ++(2.5,0)
          (10,3) node[circ]{}
        ;\end{circuitikz}\end{center}
        \caption{Full battery ECM (MF is modeled with a single Zarc element) where $R_{\text{HF}}$ and $R_{\text{MF}}$ are HF and MF resistance, respectively, and $\bar{Z}_{\text{CPE}}^{\text{xF}}$ is $\text{xF}$ CPE impedance, $\text{x}\in\left\{\text{H,M,L}\right\}$, with the coefficient $Q_{\text{xF}}$ and exponent $\phi_{\text{xF}}$.}
        \label{fig:subA_ECM_full}
    \end{subfigure}

    \vspace{0.5cm}

    \begin{subfigure}{0.8\textwidth}
        \centering
        \begin{center}\begin{circuitikz}\draw
          (0,3) to[R, l=\mbox{$R_\Sigma$}, a={$R_{\text{HF}} + R_{\text{MF}}$}] ++(2.5,0)
          (2.5,3) node[circ]{}
          (0,3) node[circ]{}
          (2.5,3) to[european resistor, l=\mbox{$Q_{LF},\phi_{LF}$}, a={$\bar{Z}_{CPE}^{LF}$}] ++(2.5,0)
          (5,3) node[circ]{}
        ;\end{circuitikz}\end{center}
        \caption{ECM seen at LF sampling where $R_\Sigma$ captures the total HF and MF resistance.}
        \label{fig:subB_ECM_at_Low_CPE}
    \end{subfigure}

    \vspace{0.5cm}

    \begin{subfigure}{0.8\textwidth}
        \centering
            \begin{center}\begin{circuitikz}\draw
      (-0.5,2.5) node[circ]{}
      (-0.5,2.5) to[R, l=\mbox{$R_\Sigma$}] ++(2,0)
      (1.5,2.5) node[circ]{}
      (1.5,2.5) to[R, l=\mbox{$R_\infty$}] ++(2,0)
      (3.5,2.5) node[circ]{}
      (3.5,3) to[short] ++(0,-1.5)
      to[short] ++(0,2)
      (3.5,3.5) to[R, l=\mbox{$R_1$}] ++(2,0)
      (3.5,1.5) to[C, l=\mbox{$C_1$}] ++(2,0)
      (5.5,3.5) to[short] ++(0,-2)
      (5.5,2.5) node[circ]{}
      (5.5,2.5) to[short] ++(0.5,0)
      (7,2.5) to[short] ++(0.5,0)
      (7.5,3.5) to[R, l=\mbox{$R_n$}] ++(2,0)
      (7.5,3.5) to[short] ++(0,-2)
      to[C, l=\mbox{$C_n$}] ++(2,0)
      (9.5,3.5) to[short] ++(0,-2)
      (9.5,2.5) to[short] ++(0.5,0)
      (7.5,2.5) node[circ]{}
      (9.5,2.5) node[circ]{}
      (10,2.5) node[circ]{}
      (6.5,2.15) node[label=$\cdots$]{};
    ;\end{circuitikz}\end{center}
        \caption{Approximated ECM with decomposed LF CPE to $n$ RC branches, with resistances $R_1,\dots,R_n$ and capacitances $C_1,\dots,C_n$ and the corrective elements $R_{\infty}$ and $C_{\infty}$}
        \label{fig:subC_ECM_at_Low_Approx_CPE}
    \end{subfigure}

    \caption{ECM and its approximations due to LF sampling.}
\end{figure}
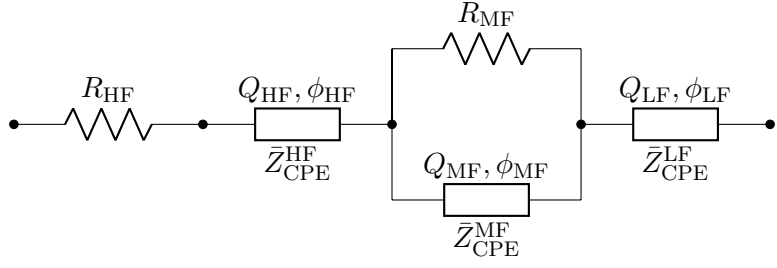
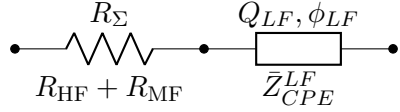
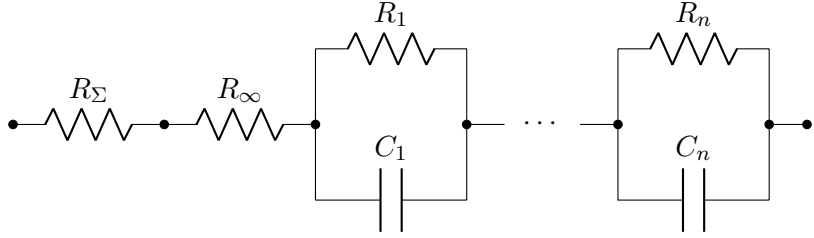

With limited BMS sampling frequency, both mid- and high-frequency behavior of the battery are unobservable. In this respect, the HF and MF CPEs of the original ECM (shown in Fig.~\ref{fig:subA_ECM_full}) are replaced with short- and open-connections between their terminals and, therefore, the ECM seen at LF sampling consists of the total resistance $R_\Sigma = R_{\text{HF}} + R_{\text{MF}}$ and the LF CPE.

Both complete and reduced circuits are suitable for the frequency domain modelling, but due to the presence of CPEs, which are fractional order elements, in the time domain, this would require writing and solving fractional order differential equations (FODE). However, for a defined frequency range, the CPE can be decomposed using the proposed method by Valsa \cite{valsa_network_2011},\cite{valsa_rc_2013} to $n$ parallel RC branches with additional corrective elements, $R_\infty$ and $C_\infty$. Fig.~1 depicts: (a) Full battery ECM, (b) ECM seen at LF where $R_\Sigma$ captures the total HF and MF resistance and (c) approximated ECM with decomposed LF CPE to $n$ RC branches.

\subsection{CPE Approximations}

According to \cite{plett_battery_2015}, the exact equivalence of CPE would require the connection of infinite RC branches, each of them having larger and larger time constants. Due to incapability to afford this in a real setting, we need to find a way to approximate the circuit with a finite number of electrical elements. This, obviously, directly limits the frequency range where the decomposed circuits approximate well the CPE behaviour. This is also the case with the model proposed in \cite{valsa_network_2011}. We recall that the reason why the authors included the corrective elements $R_\infty$ and $C_\infty$ is to compensate for the fact that, after $n$ RC branches, an infinite number of RC branches are missing for the exact equivalence.

\begin{figure*}[htb!]
\centering
\begin{subfigure}{.52\textwidth}
  \centering
  \includegraphics[width=0.9\linewidth]{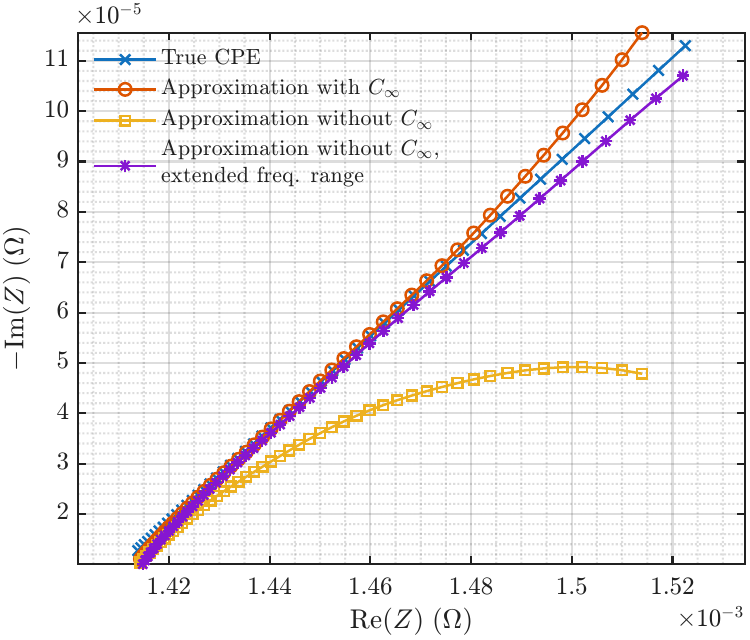}
  \caption{}
  \label{SubFig:approximations}
\end{subfigure}%
\begin{subfigure}{.48\textwidth}
  \centering
  \includegraphics[width=0.9\linewidth]{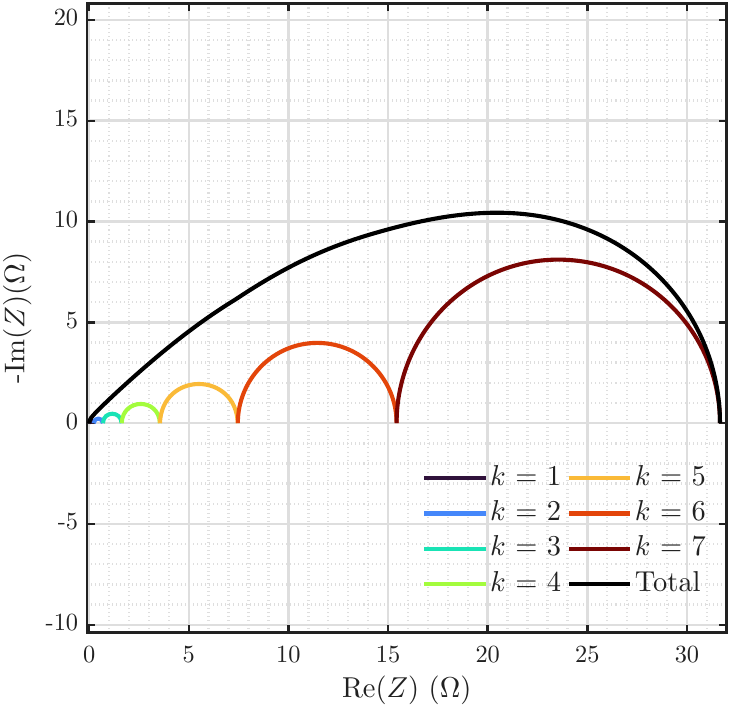}
  \caption{}
  \label{SubFig:ladder}
\end{subfigure}
\caption{
(a) Comparison of the approximated ECMs without the corrective capacitor - for both the original and the extended frequency ranges compared to the approximated ECM with the corrective capacitor and the original CPE; and (b) Nyquist plot of the decomposed CPE ECM with seven RC branches with recursive definition. Coloured curves represent the impedance contribution of individual RC branches, while the black curve shows the total impedance.
}
\label{fig:approx_and_ladder}
\end{figure*}

In the time domain, however, the usage of an ECM comprising the series capacitor $C_\infty$ can introduce initialization issues, since the initial voltage across this capacitor is undefined and largely affecting the parameters identification reliability. To avoid this problem, the model is adjusted by removing $C_\infty$. The impedance of the ideal, non-approximated CPE is taken as a reference. The theoretical locus of a CPE in the Nyquist plane is a straight line with an inclination angle $\gamma$, that depends on the CPE exponent, namely $\gamma = \frac{\pi\phi}{2}$. The CPE coefficient $Q$ determines how the impedance points are distributed along this line. As illustrated in Fig.~\ref{SubFig:approximations}, removing the corrective elements results in a Nyquist response that remains nearly linear over a certain frequency range but gradually bends toward the real axis at lower frequencies. This bending marks the deviation from the ideal CPE locus, and therefore the validity of the approximation limited to the initial, high-frequency portion of the plot. Increasing the number of RC branches extends the frequency range of validity, shifting the onset of bending toward lower frequencies and widening the linear region.

The Nyquist plot shown in Fig.~\ref{SubFig:ladder} illustrates the contribution of each RC branch and their combined effect on the total impedance. Each coloured semicircle corresponds to one parallel RC element, representing its corresponding characteristic relaxation process of the battery. As frequency decreases, successive branches dominate, shifting the response along the real axis. We recall that, the total impedance (plotted in black) is the combination of all branches and shows a smooth transition across frequency decades, forming a continuum that approximates a CPE response over a defined frequency range before gradually bending at low frequencies.

\subsection{Decomposed CPE model}\label{sec:DecomposedCPEmodel}
The complex CPE impedance of the ECM shown in Fig.~\ref{fig:subB_ECM_at_Low_CPE} and  is given by (for the sake of readability, we omit the use of LF in the superscript): 
\begin{equation}
    \bar{Z}_{\text{CPE}}(Q,\phi; \omega) = \dfrac{1}{Q(j\omega)^\phi}.
\end{equation}
For the sake of consistency with later derivations, in what follows we summarize the steps for decomposing CPE, as proposed in \cite{valsa_network_2011}.
Given the CPE coefficient $Q$ and exponent $\phi$, these steps provide the approximated (decomposed) ECM circuit parameters with $n$ parallel RC branches connected in series, where branch resistances and capacitances follow recursive laws:
\begin{equation}
    R_k = R_1 a^{k-1}\; \text{and} \; C_k = C_1 b^{k-1}, \quad k=1,\dots, n
\end{equation}
with $0<a<1$ and $0<b<1$, and the corrective resistance is defined as:
\begin{equation}\label{eq:Rinf_definition}
    R_{\infty} = R_1\dfrac{a^n}{1-a}
\end{equation}
Therefore, knowing $R_1, C_1, a$ and $b$ enables us to compute all the parameters of the decomposed CPE circuit. 
The time constants of RC branches are in decreasing order:
\begin{equation}
    \tau_{\min} = \tau_1 =R_1 C_1 > \cdots > R_n C_n = \tau_n =\tau_{\max}
\end{equation}
Obviously, for $k$-th time constant, $k>1$, $\tau_k = \tau_{\min}\cdot(ab)^{k-1}$. For characteristic frequencies of each RC branch, we have:
\begin{equation}\label{eq:f_min_and_f_max_order}
    f_{\min} = \dfrac{1}{2\pi {R_1 C_1}}  = f_{1} < \cdots <  f_{n} = \dfrac{1}{2\pi R_n C_n} = f_{\max}
\end{equation}
For $k$-th frequency, $k>1$ $f_k = \dfrac{f_{\min}}{(ab)^{k-1}}$. Minimum and maximum frequencies are related as $f_{\min} = f_{\max}\cdot(ab)^{n-1}$.
Such an approximation with a finite number of RC branches produces a ripple in impedance phase. Ratios $a$ and $b$ are related to the CPE exponent and defined as:
\begin{equation}\label{eq:a_and_b_definition}
        a  = q^\phi \quad \text{and} \quad
        ab = q \Rightarrow  b = \dfrac{q}{a}.
\end{equation}
where the coefficient $q$ is a constant and depends only on the maximum phase ripple (in radians):
\begin{equation}
    q = \dfrac{0.24}{1 + \Delta_{\phi}\cdot 180/\pi}.
\end{equation}
The approximated impedance is given by:
\begin{equation}
    \bar{Z}_{app}(\omega) = R_{\infty} + \sum_{k=1}^n\dfrac{R_k}{1+j\omega R_k C_k}
\end{equation}
and can be expressed in terms of $R_1$, $C_1$, $a$ and $b$ as:
\begin{equation}\label{eq:Z_app_sum_R1_C1_a_b}
    \bar{Z}_{app}(R_1, C_1, a, b; \omega) = R_1\dfrac{a^n}{1-a} + \sum_{k=1}^n\dfrac{a^{k-1}R_1}{1+j\omega (ab)^{k-1} R_1 C_1}
\end{equation}
To obtain the circuit parameters that approximate a CPE with coefficient $Q$, we require that: 
\begin{equation}\label{eq:Zapp_equals_Z_CPE_at_omega_avg}
 |\bar{Z}_{app}(R_1,C_1,a,b;\omega_{avg})| = |\bar{Z}_{\text{CPE}}(Q,\phi; \omega_{avg})| = \dfrac{1}{Q\cdot \omega_{avg}^\phi}
\end{equation}
where the average angular frequency is defined by:
\begin{equation}\label{eq:omega_avg_prims}
    \omega_{avg} = \left(\dfrac{a}{b}\right)^{0.25}\dfrac{1}{R_1' C_1' q^{\lceil\frac{n}{2} - 1\rceil}}.
\end{equation}
In other words, \eqref{eq:Zapp_equals_Z_CPE_at_omega_avg} ensures that, at $\omega_{avg}$, the equivalent impedance of the decomposed CPE circuit matches, in magnitude, the corresponding impedance of the exact CPE.
The values for $R_1$ and $C_1$ are obtained starting from an arbitrary resistance, $R_1'$ and capacitance, $C_1'$,  that correspond to the minimum characteristic frequency, $f_{\min}$:
\begin{equation}\label{eq:R1C1_equals_R1primC1prim}
    R_1'C_1' = \frac{1}{2 \pi f_{\min}} = R_1 C_1
\end{equation}
and then multiplying the resistance $R_1'$ and dividing the capacitance $C_1'$ by the following gain: 
\begin{equation}\label{eq:gain_definition}
    g = \dfrac{|\bar{Z}_{\text{CPE}}(Q,\phi; \omega_{avg})|}{|\bar{Z}_{app}(R_1',C_1',a,b;\omega_{avg})|} = \dfrac{1}{Q\cdot \omega_{avg}^\phi|\bar{Z}_{app}(R_1',C_1',a,b;\omega_{avg})|}.
\end{equation}
Therefore:
\begin{equation}\label{eq:gain_apply_to_R1prim_and_C1prim}
    R_1 = g R_1' \quad \text{and} \quad C_1 = \dfrac{1}{g} C_1'.
\end{equation}

\subsection{Discrete state-space representation of battery LF ECM with decomposed CPE}

\begin{figure}[htb!]
\begin{center}\begin{circuitikz}[scale=0.8, transform shape, font = \Large]\draw
  (1.5,2.5) node[circ]{}
  (3.5,2.5) node[circ]{}
  (3.5,3) to[short] ++(0,-1.5)
  to[short] ++(0,2)
  (3.5,3.5) to[R, l=\mbox{$R_1$}] ++(3,0)
  (3.5,1.5) to[C, l=\mbox{$C_1$}, v_<=$v_1$] ++(3,0)
  (6.5,3.5) to[short] ++(0,-2)
  (6.5,2.5) node[circ]{}
  (6.5,2.5) to[short] ++(0.5,0)
  (8,2.5) to[short] ++(0.5,0)
  (8.5,3.5) to[R, l=\mbox{$R_n = R_1 a^{n-1}$}] ++(3,0)
  (8.5,3.5) to[short] ++(0,-2)
  to[C, l=\mbox{$C_n = C_1 b^{n-1}$}, v_<=$v_n$] ++(3,0)
  (11.5,3.5) to[short] ++(0,-2)
  (11.5,2.5) to[short, -*, i=$i$] ++(1.5,0)
  (8.5,2.5) node[circ]{}
  (11.5,2.5) node[circ]{}
  (1.5,0) to[short] ++(11.5,0)
  to[open, v=$v$] ++(0,2.5)
  (1.5,2.5) to[battery,l_=\mbox{$E_0$}] ++(0,-2.5)
  (1.5,2.5) to[R, l=\mbox{$R_\Sigma + R_\infty$}, v_<=$v_\Sigma$] ++(2,0)
  (7.54,2.1) node[label=$\cdots$]{};
;\end{circuitikz}
\caption{Cell ECM with decomposed CPE consisting of the equivalent resistance $R_\Sigma$ (jointly capturing HF and MF ohmic behaviour), corrective element $R_{\infty}$ and $n$ RC branches with recuresively defined resistances and capacitances.}
\label{fig:ECM}\end{center}
\end{figure}
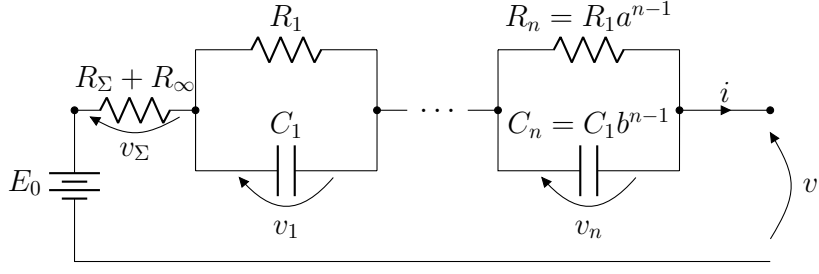

In view of considerations made in Sec.~\ref{sec:BatteryECM_at_LF} -- \ref{sec:DecomposedCPEmodel} the battery LF ECM used is the one shown in Fig.~\ref{fig:ECM}.

Writing both current and voltage Kirchoff laws for the circuit in Fig.~\ref{fig:ECM} leads to system of differential equations that can be written in a compact form as the following continuous state-space model. The voltage Kirchoff law yileds:
\begin{equation}
\label{eq:v_cell_2nd_Kirchhoff}
    v(t) = E_0 - (R_\Sigma+R_{\infty})\cdot  i - \sum_{k=1}^n v_k(t)
\end{equation}
and for the circuit current $i$ we have, for each RC branch:
\begin{equation}\label{eq:current_iR_plus_iC}
    i(t) = i_{R_k}(t) + i_{C_k}(t) \quad \forall k = 1,\dots,n,
\end{equation}
where
\begin{equation}\label{eq:currents_i_R_and_i_C}
    i_{R_k} = \dfrac{v_k(t)}{R_k}\quad\text{and}\quad  i_{C_k} = \dfrac{dv_{k}(t)}{dt} 
\end{equation}
Combining \eqref{eq:current_iR_plus_iC} and \eqref{eq:currents_i_R_and_i_C} results in:
\begin{equation}
    \dfrac{dv_k(t)}{dt} = -\dfrac{1}{R_k C_k} v_k(t) + \dfrac{i(t)}{C_k},\; k = 1,\dots,n
\end{equation}
We assume that the OCV curve is known\footnote{We assume the modeller to be provided with the standard $OCV(SOC)$ curve of the battery under study.} and discretized as a piecewise linear function. Therefore, its each segment can be expressed as: 
\begin{equation}
\label{eq:E_SOC_linear}
    E(\text{SOC}) = \alpha_l(\text{SOC}) + \beta_l(\text{SOC})\cdot\text{SOC}, \quad l = 1,...,N_{OCV}
\end{equation}
where $\alpha_l$ and $\beta_l$ are known coefficients for each of $N_{OCV}$ segments, in total. The evolution of SOC is modelled by applying Coulomb counting as:
\begin{equation}\label{eq:Coulomb_counting_SOC}
    \text{SOC}(t) = \text{SOC}_0 -\dfrac{1}{C_\text{nom}}\int_0^t i(\tau) d\tau
\end{equation}
with the nominal charge capacity (in Ah) of the battery, $C_\text{nom}$ at a given discharge rate and temperature.
Let us consider a range of SOC for which the ECM parameters can be considered as constant, and the OCV curve is linear, with coefficients $\alpha$ and $\beta$. Then, by combining \eqref{eq:v_cell_2nd_Kirchhoff}-\eqref{eq:Coulomb_counting_SOC}, we can write the continuous state-space model as:
\begin{equation}\label{eq:ss_continuous}
    \begin{aligned}
        \dfrac{d}{dt}{\mathbf{x}}(t) &= \mathbf{A}_c(\boldsymbol{\Theta}) \mathbf{x}(t) + \mathbf{B}_c(\boldsymbol{\Theta}) \mathbf{u}(t) \\
        {y}(t) &= \mathbf{C}_c(\boldsymbol{\Theta}) \mathbf{x}(t) + \mathbf{D}_c(\boldsymbol{\Theta}) \mathbf{u}(t)\\
    \end{aligned}
\end{equation}
where the system state is:
\begin{equation}\label{eq:vector_x}
    \mathbf{x}(t) =
        \begin{bmatrix}
        v_{1}(t), \dots, v_n(t), \text{SOC}(t)
        \end{bmatrix}^\top
\end{equation}
and the input and output time-domain signals are

\begin{equation}\label{eq:vectors_u_and_y}
   \mathbf{u}(t) = [i(t), 1]^\top \;\; \text{and} \;\; y(t) = v(t)
\end{equation}

\noindent Matrices\footnote{\label{fn:matrices_cont_and_disc}With indices $c$ and $d$ we distinguish between continuous and discrete formulation.} $\mathbf{A}_c$, $\mathbf{B}_c$, $\mathbf{C}_c$ and $\mathbf{D}_c$ are dependent on the  parameter vector
\begin{equation}
    \boldsymbol{\Theta} = \left[R_\Sigma, R_1, C_1, a, b\right]^\top
\end{equation}
and derived as follows:

\begin{equation}\label{eq:matrix_Ac}
    \begin{aligned}
            \mathbf{A}_c(\boldsymbol{\Theta}) &=
            \begin{bmatrix} 
                -\tfrac{1}{R_1 C_1} & 0                      & \dots  &                     0 & 0\\
                0                   & -\tfrac{1}{R_2 C_2}    & \dots  &                     0 & 0\\
                \vdots              & \vdots                 & \ddots &    \vdots             & 0\\
                0                   &0                       & \dots  &   -\tfrac{1}{R_n C_n} & 0\\
                0                   &0                       &\dots   &                     0 & 0\\
            \end{bmatrix}  = -\tfrac{1}{R_1 C_1}\begin{bmatrix}
                1 & 0                      & \dots & 0 &             0 \\
                0                   & \tfrac{1}{ab}    & \dots & 0 & 0 \\
                \vdots              & \vdots           & \ddots & \vdots  & \vdots \\
                0                   &0                        & \dots &   \tfrac{1}{(ab)^{n-1}} & 0 \\
                0                   &0                        &\dots  & 0 & 0 \\
            \end{bmatrix}
    \end{aligned}
\end{equation}

\begin{equation}\label{eq:matrix_Bc}
    \mathbf{B}_c(\boldsymbol{\Theta}) =
        -\begin{bmatrix}
        \tfrac{1}{C_1}  & 0\\
        \vdots & \vdots \\
        \tfrac{1}{C_n} & 0 \\
        -\tfrac{1}{C_\text{nom}} & 0\\
        \end{bmatrix}
        =
        -\begin{bmatrix}
        \tfrac{1}{C_1}  & 0 \\
        \vdots  & \vdots\\
        \tfrac{1}{b^{n-1}C_1} & 0 \\
        -\tfrac{1}{C_\text{nom}} & 0
        \end{bmatrix}
\end{equation}

\begin{equation}\label{eq:matrix_Cc}
            \mathbf{C}_c(\boldsymbol{\Theta}) =
       {[
            \underbrace{-1,\,-1,\,-1,\,\dots,\,-1}_{n}, \; \beta
            ]}
\end{equation}

\begin{equation}\label{eq:matrix_Dc}
    \mathbf{D}_c(\boldsymbol{\Theta}) =
    \begin{bmatrix}
                -R_\Sigma-R_{\infty}, \; \alpha
            \end{bmatrix} =
            \begin{bmatrix}
                -R_\Sigma-R_1\dfrac{a^n}{1-a}, \; \alpha
            \end{bmatrix}
\end{equation}
Since the measured input and output signals are, in practice, discrete, we use a discrete state-space model, which also simplifies the estimation model via finite differences. The equations of discretized state-space model become:
\begin{equation}\label{eq:ss_discrete}
    \begin{aligned}
        {\mathbf{x}}_{k+1}(\boldsymbol{\Theta}) &= \mathbf{A}_d(\boldsymbol{\Theta}) \mathbf{x}_k + \mathbf{B}_d(\boldsymbol{\Theta})\mathbf{u}_k \\
        y_k(\boldsymbol{\Theta})    &= \mathbf{C}_d(\boldsymbol{\Theta}) \mathbf{x}_k + \mathbf{D}_d(\boldsymbol{\Theta})\mathbf{u}_k \\
    \end{aligned}
\end{equation}
where the discretized system state, input and output signals are
\begin{equation}\label{eq:vector_x}
    \mathbf{x}_k =
        \begin{bmatrix}
        v_{1}(k), \dots, v_n(k), \text{SOC}_k
        \end{bmatrix}^\top
\end{equation}
\begin{equation}\label{eq:vectors_u_and_y}
   \mathbf{u}_k = [i_k, 1]^\top \;\; \text{and} \;\; y_k = v(k)
\end{equation}
and matrices\footref{fn:matrices_cont_and_disc} derived from the continuous formulation as:
\begin{subequations}\label{eq:matrices_ABCD_d}
    \begin{align}
        \mathbf{A}_d(\boldsymbol{\Theta}) & = \mathbf{I}+ T_s\cdot\mathbf{A}_c(\boldsymbol{\Theta})\\
        \mathbf{B}_d(\boldsymbol{\Theta}) & = T_s\cdot\mathbf{B}_c(\boldsymbol{\Theta})\\
        \mathbf{C}_d(\boldsymbol{\Theta}) & =\mathbf{C}_c(\boldsymbol{\Theta})\\
        \mathbf{D}_d(\boldsymbol{\Theta}) & =\mathbf{D}_c(\boldsymbol{\Theta})
    \end{align}
\end{subequations}
With $\mathbf{I}$ we denote $(n+1)$-dimensional identity matrix and $T_s$ is the sampling time.

Note that Coulomb counting as a simplified SOC estimation method is here included directly in the state-space model, however, other SOC estimation strategies can be also applied. After SOC is estimated, the reconstructed OCV (knowing the OCV-SOC curve) contribution can be removed from the measured voltage and used as the output. In that case, state-space matrices would be reduced by removing the last row and column for $\mathbf{A}$ and $\mathbf{B}$ and last elements from vectors $\mathbf{C}$ and $\mathbf{D}$. Note that inclusion of other SOC estimator methods is beyond the scope of this paper.

\subsection{Parameter estimation problem}
We suppose that we collect discrete measurements of the voltage and current of the battery cell via a BMS, with sampling time $T_s$ that is bounded to several hundreds of ms (a condition usually imposed by existing BMSs). As already discussed, this forces the modeler to observe only the total resistance $R_\Sigma$ and LF parameters. The goal is to estimate the parameter set $\boldsymbol{p} = \left[R_\Sigma, Q, \phi\right]^\top$. This will be done by first estimating the parameters $\boldsymbol{\Theta}$ of the ECM, which includes the decomposed CPE and, then, reconstructing the CPE elements following the decomposition steps in reverse order.

The number of branches to approximate the CPE depends on the minimum and maximum frequency of interest \cite{valsa_network_2011}:
\begin{equation}\label{eq:num_of_branches}
  n = \left\lceil\dfrac{\ln f_{\min}-\ln f_{\max}}{\ln q}\right\rceil
\end{equation}
First, to satisfy the Nyquist-Shannon sampling theorem, the maximum frequency $f_{\max}$ is at most $f_{\max} = f_s/2$, where $f_s = 1/T_s$ is the sampling frequency. Additionally, to provide a stable response of the discrete state-space model introduced in the previous section, integrated with forward Euler, the theoretical limit is $f_s/\pi$, which is derived in the Appendix \ref{app:Appendix_Stability_limit}. In practice, depending on the excitation signal and chosen number of RC branches, the maximum estimated frequency (corresponding to the lowest estimated time constant), can be lower than this value. This introduces another dimension in the parameter estimation problem since we do not know in advance the maximum frequency that corresponds to the decomposition. Therefore, we allow:
\begin{equation}
    f_{\max}^{\downarrow} \leqslant f_{\max}\leqslant f_{\max}^{\uparrow}
\end{equation}

From the observability point of view, $f_{\min}$ in \eqref{eq:num_of_branches} is at least equal to the frequency resolution, given the number of samples \cite{valsa_network_2011}:
\begin{equation}
    f_{\min} \geqslant \dfrac{f_s}{N}
\end{equation}
whereas the maximum frequency is bounded by:
\begin{equation}
    f_{\max} \leqslant \dfrac{f_s}{\pi}
\end{equation}
Therefore, the maximum number of branches would be:
\begin{equation}
   \left\lceil\dfrac{\ln f_s/N-\ln f_s/\pi}{\ln q} \right\rceil
\end{equation}
assuming that each of time constants has an arbitrary value. However, under the assumption that time constants follow geometric series, increasing the number of branches beyond this limit (recall that a true CPE ECM has infinite number of RC branches), actually has a positive effect to the estimation results: it will allow us to obtain better estimates for observable RC branches (with lowest time constants) regardless the fact that some of the branches contributions will not be observable (due to the presence of noise) given the excitation.

We formulate the optimization problem to estimate the parameters $\boldsymbol{\Theta}$ given $N$ samples of measured input and output signals, $\tilde{u}_k$ and $\tilde{y}_k$, $k=0,\dots, N-1$.

For a fixed number of RC branches and assigned values to the decomposition constants $q$, $\Delta_{\phi}$, the vector of unknown parameters $\boldsymbol{\Theta}$ is augmented with $f_{\max}$ as follows:
\begin{equation}\label{eq:theta_aug_with_fmax}
    \boldsymbol{\Theta}' = \begin{bmatrix}
        \boldsymbol{\Theta}\\
        f_{\max}
    \end{bmatrix} = \left[R_\Sigma, R_1, C_1, a, b, f_{\max}\right]^\top = \left[\Theta_1', \Theta_2', \Theta_3', \Theta_4', \Theta_5', \Theta_6'\right]^\top
\end{equation}
Knowing their values enables us to derive all the ECM parameters and completely reconstruct its response. The parameters are estimated by solving the following least squares problem:
\begin{subequations}\label{eq:optimization_prob}
    \begin{align}
        \min_{\boldsymbol{\Theta}'} \quad  & \left\|\mathbf{\tilde{y}}-\mathbf{y}(\boldsymbol{\Theta}')\right\|_2^2\label{eq:LS_objective} \\[1.5ex]
        \text{subject to}    \quad & \eqref{eq:matrix_Ac}-\eqref{eq:matrices_ABCD_d} \label{eq:matrix_constraints}, \eqref{eq:theta_aug_with_fmax}\\
                                   & ab = q \label{eq:ab_eq_q} \\
                                   & \dfrac{1}{2\pi R_1 C_1}=f_{\min} \label{eq:f_min_R1_C1}\\
                                   & f_{\min} = f_{\max}q^{n-1} \label{eq:f_min_f_max_qn}\\
        & \boldsymbol{\Theta}'_{\min} \leqslant \boldsymbol{\Theta}' \leqslant \boldsymbol{\Theta}'_{\max} \label{eq:theta_min_max}
    \end{align}
\end{subequations}
where we minimise the distance between the collection of samples of measured output voltage and the corresponding model, expressed as a function of unknown parameters, when a cell is excited with a known (measured) current\footnote{This is a classical ordinary least squares problem assuming a homoscedastic noise model. In case the modeler has access to a more precise noise model, the objective given by \eqref{eq:LS_objective} can be modified by introducing weighting through a covariance matrix of the measurements.}. Considering $N$ samples (measurements), the measured and modelled output vectors are $\mathbf{\tilde{y}} = \left[\tilde{y}_0,\dots,\tilde{y}_{N-1}\right]^\top$ and $\mathbf{y(\boldsymbol{\Theta}')} = \left[{y}_0(\boldsymbol{\Theta}'),\dots,y_{N-1}(\boldsymbol{\Theta}')\right]^\top$, respectively. The equality constraint \eqref{eq:ab_eq_q} arises directly from \eqref{eq:a_and_b_definition} used for the CPE decomposition. The lower and upper bounds on each parameters are given by the last inequality constraint. How to set their values, $\boldsymbol{\Theta}'_{\min}$ and $\boldsymbol{\Theta}'_{\max}$ is described in part \ref{sec:parameter_bounds}.

\subsubsection{Parameter Bounds}\label{sec:parameter_bounds}
\begin{figure}[htb!]
    \centering
    \includegraphics[scale = 0.48]{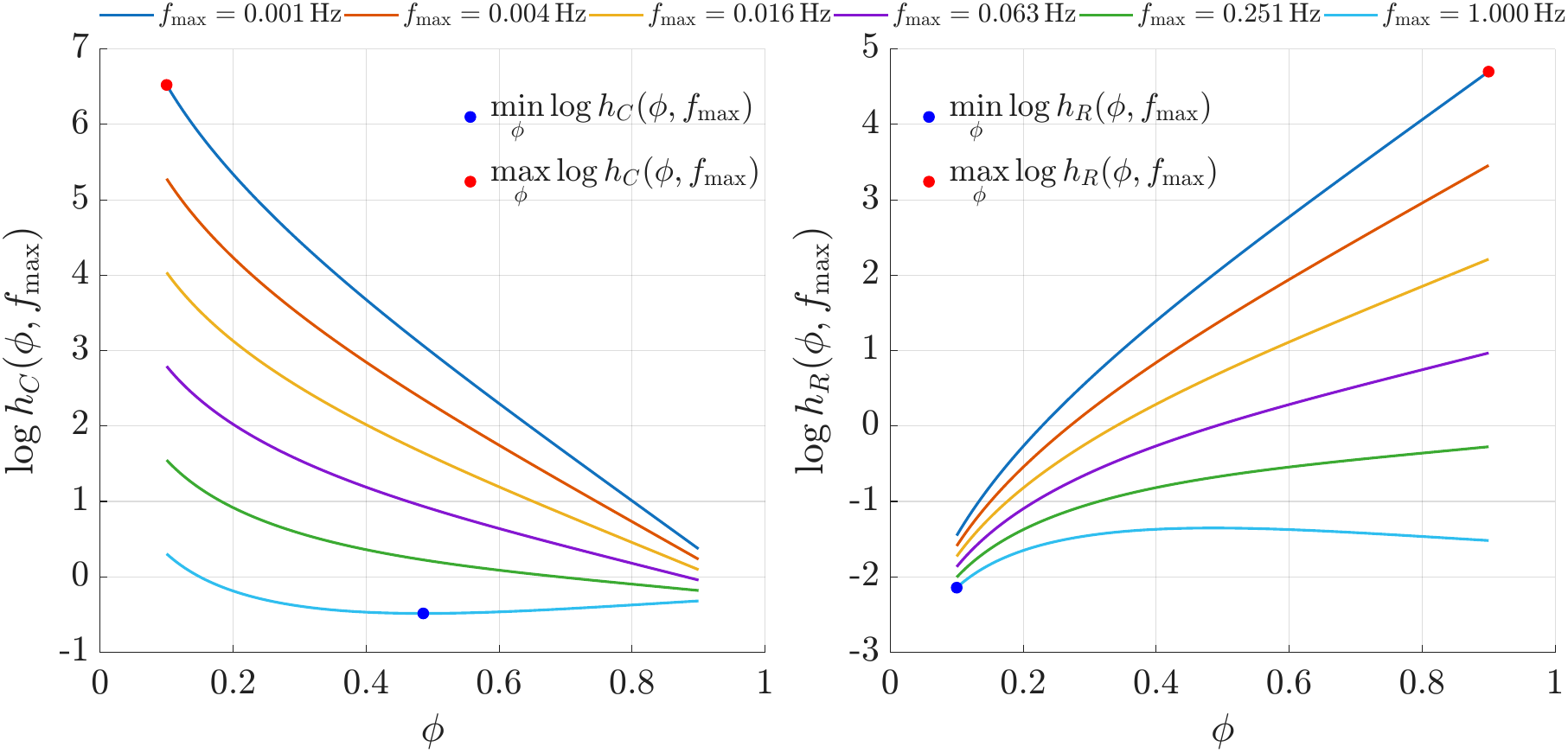}
    \caption{Family of curves showing the dependency of $h_C(\phi,f_{\max})$ and $h_R(\phi,f_{\max})$ on the LF CPE exponent $\phi$, for different values of the maximum frequency $f_{\text{max}}$ (in logarithmic scale).}
    \label{fig:family_of_curves}
\end{figure}

To help the solver find the optimal solution more efficiently, we here describe how to impose lower and upper limits, $\boldsymbol{\Theta}'_{\min}$ and $\boldsymbol{\Theta}'_{\max}$ on each parameter in $\boldsymbol{\Theta}'$ deriving them from chosen lower and upper bounds on $\boldsymbol{p}$. Namely, we first set the bounds on every element of $\boldsymbol{p}$:

\begin{equation}
    \boldsymbol{p}_{\min}=
    \begin{bmatrix}
        R_{\Sigma}^{\min}\\
        Q^{\min}\\
        \phi^{\min}
    \end{bmatrix}
    \leqslant
    \begin{bmatrix}
        R_\Sigma\\
        Q\\
        \phi        
    \end{bmatrix}
    \leqslant
    \begin{bmatrix}
    R_{\Sigma}^{\max}\\
    Q^{\max}\\
    \phi^{\max}
\end{bmatrix}
    =
    \boldsymbol{p}_{\max}
\end{equation}
and translate them to lower and upper bounds on each of parameter of $\boldsymbol{\Theta}'$. First, the parameters $a$ and $b$ depend only on $\phi$ and, since $q < 1$, for a reasonable small $\Delta_\phi$, we have:
\begin{equation}\label{eq:a_min_a_max}
    a^{\min} = q^{\phi^{\max}} \quad \text{and} \quad a^{\max} = q^{\phi^{\min}}
\end{equation}
Consequently,
\begin{equation}\label{eq:b_min_b_max}
    b^{\min} = q/a^{\max} \quad \text{and} \quad b^{\max} = q/a^{\min}
\end{equation}
The bounds on $R_1$ and $C_1$, for a fixed number of RC branches, say $n$, are imposed by analyzing the gain $g$ that depends on both $Q$ and $\phi$. Let us first derive the bounds on $C_1$. Since in \eqref{eq:R1C1_equals_R1primC1prim}--\eqref{eq:gain_apply_to_R1prim_and_C1prim} we can start from arbitrary $R_1'$ and $C_1'$, here we consider $C_1' = 1~\text{F}$ and $R_1'=2\pi f_{\min} = 2\pi f_{\max}q^{n-1}$, for which $C_1$ is then inversely proportional to the gain: 
\begin{equation}
    C_1\sim \frac{1}{g} = {Q \cdot\omega_{avg}^{\phi}} |\bar{Z}_{app}(R_1',C_1',a,b;\omega_{avg})|\Bigg|_{\substack{
R_1' = 2\pi f_{\max} q^{n-1} \\
C_1' = 1
}}
\end{equation}
Therefore, the higher the coefficient, $Q$, the higher the $C_1$ value. However, the remaining product depends on $\phi$ and $f_{\max}$ and it is not, in general, monotonic function of $\phi$.
We define it as:
\begin{equation}
h_C(\phi,f_{\max}) = \omega_{avg}^{\phi}
\left| \bar{Z}_{app}(R_1', C_1', a, b; \omega_{avg}) \right|
\Bigg|_{\substack{
R_1' = 2\pi f_{\max} q^{n-1} \\
C_1' = 1
}}
\end{equation}
and numerically find the minimum and maximum values:
\begin{equation}
    h_C(\phi,f_{\max}):\quad (\phi,f_{\max}) \in \left[\phi^{\min}, \phi^{\max}\right] \times [f_{\max}^\downarrow, f_{\max}^\uparrow]
\end{equation}
Therefore, the bounds of $C_1$ can be found numerically from the family of curves (see Fig.~\ref{fig:family_of_curves}) as:
\begin{subequations}\label{eq:C_min_C_max}
    \begin{align}
        C_1^{\min} &= Q^{\min} \min_{\phi} h_C(\phi, f_{\max}),\quad  f_{\max}\in[f_{\max}^\downarrow, f_{\max}^\uparrow]\\
        C_1^{\max} &= Q^{\max} \max_{\phi} h_C(\phi, f_{\max}),\quad  f_{\max}\in[f_{\max}^\downarrow, f_{\max}^\uparrow]
    \end{align}
\end{subequations}

Similarly, to find bounds for $R_1$, if we consider $R_1' = 1~\Omega$ and $C_1'=2\pi f_{\max}q^{n-1}$, then:
\begin{equation}
R_1 \sim g = {Q \cdot \omega_{avg}^{\phi}}
\left| \bar{Z}_{app}(R_1', C_1', a, b; \omega_{avg}) \right|
\Bigg|_{\substack{
R_1' = 1 \\
C_1' = 2\pi f_{\max} q^{n-1}
}}
\end{equation}
and define:
\begin{equation}
    h_R(\phi,f_{\max}) =\omega_{avg}^{\phi} |\bar{Z}_{app}(R_1',C_1',a,b;\omega_{avg})|\Bigg|_{\substack{
R_1' = 1 \\
C_1' = 2\pi f_{\max} q^{n-1}
}}
\end{equation}
the bounds of $R_1$ can be found numerically from the family of curves (see Fig.~\ref{fig:family_of_curves}) as:
\begin{subequations}\label{eq:R_min_R_max}
    \begin{align}
        R_1^{\min} &= \dfrac{1}{Q^{\max}} \min_{\phi} \dfrac{1}{h_R(\phi, f_{\max})},\quad  f_{\max}\in[f_{\max}^\downarrow, f_{\max}^\uparrow]\\
        R_1^{\max} &= \dfrac{1}{Q^{\max}} \max_{\phi} \dfrac{1}{h_R(\phi, f_{\max})},\quad f_{\max}\in[f_{\max}^\downarrow, f_{\max}^\uparrow]
    \end{align}
\end{subequations}
Fig.~\ref{fig:family_of_curves} depicts a family of curves $h_C(\phi,f_{\max})$ and $h_R(\phi,f_{\max})$ for $\phi\in[0.1,0.9]$ and six different $f_{\max}$ values, logarithmically spaced from 0.001 Hz to 1 Hz, and the minimum and maximum attained values used to compute $C_1^{\min}$, $R_1^{\min}$, $C_1^{\max}$, and $R_1^{\max}$.

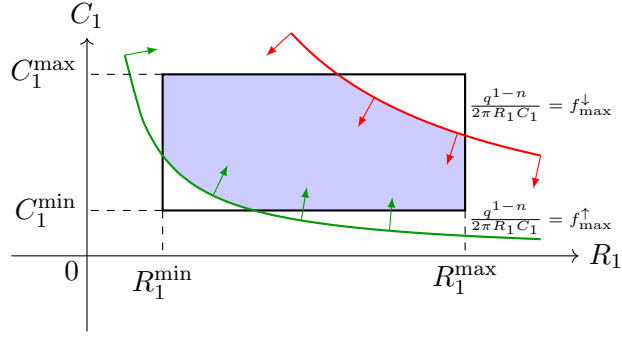
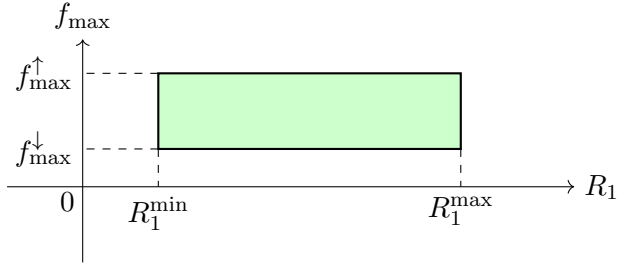
\begin{figure*}[htb!]
\centering
\begin{subfigure}{.9\textwidth}
  \centering
   \begin{tikzpicture}[scale = 1]

    \def\Rmin{1}
    \def\Rmax{5}
    \def\Cmin{0.6}
    \def\Cmax{2.4}

    \def\fA{0.02}
    \def\fC{0.12}

    \draw[->] (-1,0) -- (\Rmax*1.3,0) node[right] {$R_1$};
    \draw[->] (0,-1) -- (0,\Cmax*1.2) node[above] {$C_1$};

    \begin{scope}
        \clip (\Rmin,\Cmin) rectangle (\Rmax,\Cmax);

        \fill[blue!20, domain=\Rmin:\Rmax, variable=\R]
            plot ({\R}, {1/(2*pi*\fA*\R)})
            -- plot[domain=\Rmax:\Rmin] ({\R}, {1/(2*pi*\fC*\R)})
            -- cycle;
    \end{scope}

    \draw[thick] (\Rmin,\Cmin) rectangle (\Rmax,\Cmax);

    \draw[red, domain=2.7*\Rmin:1.2*\Rmax, thick, smooth,  variable=\R]
        plot ({\R}, {1/(2*pi*\fA*\R)});
    \node at (-0.2,-0.2) {$0$};
    
    \draw[green!60!black, domain=0.5*\Rmin:1.2*\Rmax, thick, smooth, variable=\R]
        plot ({\R}, {1/(2*pi*\fC*\R)});

    \node at (6,0.5) {\tiny $\tfrac{q^{1-n}}{2\pi R_1 C_1 }=f_{\max}^\uparrow$};
    \node at (6,2)     {\tiny $\tfrac{q^{1-n}}{2\pi R_1 C_1 }=f_{\max}^\downarrow$};

    \draw[dashed] (\Rmin,\Cmin) -- (\Rmin,0);
    \draw[dashed] (\Rmax,\Cmin) -- (\Rmax,0);
    \draw[dashed] (\Rmin,\Cmin) -- (0,\Cmin);
    \draw[dashed] (\Rmin,\Cmax) -- (0,\Cmax);

    \node[below] at (\Rmin,0) {$R_1^{\min}$};
    \node[below] at (\Rmax,0) {$R_1^{\max}$};
    \node[left] at (0,\Cmin) {$C_1^{\min}$};
    \node[left] at (0,\Cmax) {$C_1^{\max}$};

\def\N{3}
\def\scale{0.45}

\foreach \i in {0,...,\N} {
    \pgfmathsetmacro{\R}{2.7*\Rmin + (1.2*\Rmax-2.7*\Rmin)*\i/\N}
    \pgfmathsetmacro{\C}{1/(2*pi*\fA*\R)}

    \pgfmathsetmacro{\nx}{1/(\R*\R*\C)}
    \pgfmathsetmacro{\ny}{1/(\R*\C*\C)}

    \pgfmathsetmacro{\norm}{sqrt(\nx*\nx+\ny*\ny)}
    \pgfmathsetmacro{\dx}{-\scale*\nx/\norm} 
    \pgfmathsetmacro{\dy}{-\scale*\ny/\norm}

    \draw[-latex, red] (\R,\C) -- ++(\dx,\dy);
}

\foreach \i in {0,...,\N} {
    \pgfmathsetmacro{\R}{0.5*\Rmin + (0.8*\Rmax-0.5*\Rmin)*\i/\N}
    \pgfmathsetmacro{\C}{1/(2*pi*\fC*\R)}

    \pgfmathsetmacro{\nx}{1/(\R*\R*\C)}
    \pgfmathsetmacro{\ny}{1/(\R*\C*\C)}

    \pgfmathsetmacro{\norm}{sqrt(\nx*\nx+\ny*\ny)}
    \pgfmathsetmacro{\dx}{\scale*\nx/\norm}
    \pgfmathsetmacro{\dy}{\scale*\ny/\norm}

    \draw[-latex, green!60!black] (\R,\C) -- ++(\dx,\dy);
}

\end{tikzpicture}
  \caption{Feasible region for $(R_1,C_1)$ - non-convex set.}
  \label{SubFig:R1C1_nonconvex}
\end{subfigure}%
\\
\begin{subfigure}{.9\textwidth}
  \centering
      \begin{tikzpicture}

    \def\Rmin{1}
    \def\Rmax{5}
    \def\fmaxdown{0.5}
    \def\fmaxup{1.5}

    \def\fA{0.02}
    \def\fC{0.12}

    \draw[->] (-1,0) -- (\Rmax*1.3,0) node[right] {$R_1$};
    \draw[->] (0,-1) -- (0,\fmaxup*1.3) node[above] {$f_{\max}$};


    \draw[thick, fill = green!20] (\Rmin,\fmaxdown) rectangle (\Rmax,\fmaxup);

    \node at (-0.2,-0.2) {$0$};
    

    \draw[dashed] (\Rmin,\fmaxdown) -- (\Rmin,0);
    \draw[dashed] (\Rmax,\fmaxdown) -- (\Rmax,0);
    \draw[dashed] (\Rmin,\fmaxdown) -- (0,\fmaxdown);
    \draw[dashed] (\Rmin,\fmaxup) -- (0,\fmaxup);

    \node[below] at (\Rmin,0) {$R_1^{\min}$};
    \node[below] at (\Rmax,0) {$R_1^{\max}$};
    \node[left] at (0,\fmaxdown) {$f_{\max}^{\downarrow}$};
    \node[left] at (0,\fmaxup) {$f_{\max}^{\uparrow}$};

\end{tikzpicture}
  \caption{Feasible region for $(R_1,f_{\max})$ - convex set.}
  \label{SubFig:R1fmax_convex}
\end{subfigure}
\caption{Feasible regions depending on the problem formulation.}
\label{fig:R1C1vsR1fmax}
\end{figure*}

\noindent Due to the fact that $f_{\max} = \frac{1}{2\pi R_1 C_1 q^{n-1}}$ and $f_{\max}\in[f_{\max}^{\downarrow}, f_{\max}^{\uparrow}]$, additional inequalities restrict $(R_1,C_1)$ feasible set:
\begin{equation}
   f_{\max}^{\downarrow} \leqslant  \frac{1}{2\pi R_1 C_1 q^{n-1}}\leqslant f_{\max}^{\uparrow}
\end{equation}
which, together with $ R_1^{\min}\leqslant R_1\leqslant R_1^{\max}$ and $ C_1^{\min}\leqslant C_1\leqslant C_1^{\min}$, yields, in general, a non-convex set as shown in Fig.~\ref{SubFig:R1C1_nonconvex}. However, if instead we consider the feasible set for $(R_1, f_{\max})$, it is convex (see Fig.~\ref{SubFig:R1fmax_convex}). Obtaining any feasible point within this convex set provides a solution for $C_1$ as a consequence of \eqref{eq:f_min_R1_C1} and \eqref{eq:f_min_f_max_qn}. This motivates the reformulation of the optimization problem.

\subsubsection{Reduced Formulation}
Equality constraints in \eqref{eq:optimization_prob} enables us to reduce the number of decision variables (parameters). Namely:
\begin{subequations}
    \begin{align}
        \eqref{eq:ab_eq_q} & \Longrightarrow b = q/a\\
        \eqref{eq:f_min_R1_C1}, \eqref{eq:f_min_f_max_qn} & \Longrightarrow C_1 = \dfrac{1}{2\pi f_{\max} q^{n-1} R_1}
    \end{align}
\end{subequations}
This allows us to write the parameter vector $\boldsymbol{\Theta}'$ as: 
\begin{equation}\label{eq:Theta_prim}
    \boldsymbol{\Theta}' = [\Theta'_1, \Theta'_2, \psi_1(\Theta'_1, \Theta'_2, \Theta'_4, \Theta'_6), \Theta'_4, \psi_2(\Theta'_1, \Theta'_2, \Theta'_4, \Theta'_6), \Theta'_6]^\top
\end{equation}
with 
\begin{equation}
 \psi_1(\boldsymbol{\Theta}') = \dfrac{q^{1-n}}{2\pi \Theta'_2 \Theta'_6}\quad
\text{and} \quad \psi_2(\boldsymbol{\Theta}') = \dfrac{q}{\Theta'_4}
\end{equation}
 and therefore
we can define:
\begin{equation}\label{eq:theta_as_reduced_Theta_prim}
    \boldsymbol{\theta} = [\Theta'_1, \Theta'_2, \Theta'_4, \Theta'_6]^{\top} = [R_{\Sigma}, R_1, a, f_{\max}]^{\top}
\end{equation}
and therefore, $\boldsymbol{\Theta}'$ can be expressed as a function of $\boldsymbol{\theta}$.
Therefore, the original optimization problem can be written as follows:
\begin{subequations}\label{eq:optimization_prob_reform}
    \begin{align}
        \min_{\boldsymbol{\theta}} \quad  & \left\|\mathbf{\tilde{y}}-\mathbf{y}(\boldsymbol{\Theta}'(\boldsymbol{\theta}))\right\|_2^2 \\[1.5ex]
        \text{subject to}    \quad & \eqref{eq:matrix_Ac}-\eqref{eq:matrices_ABCD_d} \label{eq:matrix_constraints}\\
        & \eqref{eq:Theta_prim}-\eqref{eq:theta_as_reduced_Theta_prim}\\
        & \boldsymbol{\theta}_{\min} \leqslant \boldsymbol{\theta} \leqslant \boldsymbol{\theta}_{\max} \label{eq:theta_min_max}
    \end{align}
\end{subequations}
Comparing to the previously formulated problem given by \eqref{eq:optimization_prob}, where both the objective and feasible set are non-convex, the reformulated problem \eqref{eq:optimization_prob_reform} is still non-convex, however only due to the non-convexity in the objective with, instead, a convex solution space.

\subsubsection{Parameter Initialization}\label{sec:params_initialization}
As observed before, the problem \eqref{eq:optimization_prob_reform} is a non-convex one and, therefore, requires an initial guess if solved using a standard gradient-descent approach. 
In this case it is easier to set initial values for parameters $\boldsymbol{p}^0 = \left[R_\Sigma^0, Q^0, \phi^0\right]^\top$ and perform the decomposition steps described in Sec.~\ref{sec:DecomposedCPEmodel} for $CPE(Q^0, \phi^0)$ for pre-defined number of branches, $n$ and the initial maximum frequency $f_{\max}^0$.
This directly provides the initial values for $ R_1^0 $ and $ a^0 $. The initial values for parameters $\boldsymbol{p}^0$ can be assigned based on typical estimates from EIS measurements at LF. If these values are not available, the initial CPE parameters can be randomly chosen within specified lower and upper bounds.

The resistance $R_\Sigma$ is initialized by solving a least-square problem:
\begin{equation}\label{eq:R_sigma_0_LS}
    R_\Sigma^0 = \text{arg}\min_{R_\Sigma} \|R_\Sigma \tilde{\mathbf{i}} - \tilde{\mathbf{v}}\|_2^2
\end{equation}
i.e., by neglecting all the RC branches.
Therefore, the initial parameter vector is:
\begin{equation}
    \boldsymbol{\theta}^0 = \left[R_\Sigma^0, R_1^0, a^0, f_{\max}^0\right]^\top
\end{equation}
These values represent reasonable starting points to initiate the fitting process.

\subsubsection{CPE parameter reconstruction}

Now we suppose that $\boldsymbol{\hat{\theta}} = \left[\hat{R}_\Sigma, \hat{R}_1, \hat{a}, \hat{f}_{\max}\right]^\top$ is the minimizer of the optimization problem \eqref{eq:optimization_prob}, for a fixed number of branches, $n$.
Then, the estimated LF CPE parameters, $\hat{Q}$ and $\hat{\phi}$, are reconstructed as follows. From \eqref{eq:a_and_b_definition}, the phase angle is computed directly as 
    \begin{equation}\label{eq:reconstruct_phi_LF}
    \hat{\phi} = \log_{q} \hat{a}.
\end{equation}
The capacitance and its corresponding geometric series coefficient are calculated as:
\begin{equation}
    \hat{C}_1 = \dfrac{q^{1-n}}{2\pi \hat{R}_1 \hat{f}_{\max}} \quad \text{and} \quad \hat{b} = \dfrac{q}{\hat{a}}
\end{equation}
Having the decomposed parameters and the number of branches, we can reconstruct the impedance of the decomposed circuit, for an arbitrary range of frequencies, as:
\begin{equation}
     \begin{aligned}
         \bar{Z}_{app}(\hat{R}_1, \hat{C}_1, \hat{a}, \hat{b}; \hat{\omega}_{avg}) = \hat{R}_1\dfrac{\hat{a}^n}{1-\hat{a}} + \sum_{k=1}^n\dfrac{\hat{R}_1\hat{a}^{k-1}}{1+j\hat{\omega}_{avg}\hat{R}_1\hat{C}_1(\hat{a}\hat{b})^{k-1}}
    \end{aligned}
\end{equation}
where the estimated average frequency is:
\begin{equation}
    \hat{\omega}_{avg} = \left(\dfrac{\hat{a}}{\hat{b}}\right)^{0.25}\dfrac{1}{\hat{R}_1 
    \hat{C}_1 q^{\lceil\frac{n}{2} - 1\rceil}}
\end{equation}
Using \eqref{eq:Zapp_equals_Z_CPE_at_omega_avg} we can reconstruct the CPE coefficient:
\begin{equation}
    \hat{Q} = \dfrac{1}{\hat{\omega}_{avg}^{\hat{\phi}}\cdot|\bar{Z}_{app}(\hat{R}_1, \hat{C}_1, \hat{a}, \hat{b}; \hat{\omega}_{avg})|},
\end{equation}
After this step, all the parameters $\hat{\boldsymbol{p}} = [\hat{R}_{\Sigma}, \hat{Q}, \hat{\phi}]^\top$ are reconstructed.

\subsubsection{Estimation Procedure}

The estimation consists of solving the least square (LS) problem, given by \eqref{eq:optimization_prob_reform}, for different numbers of RC branches, since the observable frequency range is not exactly known. The procedure includes the initialisation of the parameters and setting their lower and upper bounds. Starting from ECM with one RC branch, we incrementally increase the number of branches until the estimated parameters no longer improve, or the maximum set number of RC branches is reached. This, together with the limit on maximum number of RC branches, is used as a criterion to terminate the estimation procedure.

\algdef{SE}[DOWHILE]{Do}{doWhile}{\algorithmicdo}[1]{\algorithmicwhile\ #1}%

\begin{algorithm}
\caption{Estimation Procedure}
\label{alg:Estimation_proc}
\begin{algorithmic}[1]
\Procedure{Estimate LF ECM Parameters}{$\mathbf{\tilde{v}}, \mathbf{\tilde{i}}$}
    \State Initialize $f_{\max}, q, \Delta_{\phi}, \boldsymbol{p}^0, \boldsymbol{p}_{\min}, \boldsymbol{p}_{\max} $, $\Delta\gets\infty$, $n\gets1$, $\varepsilon$, $n_{RC}^{\max}$
    \State Initialize $R_{\Sigma}^0$ \eqref{eq:R_sigma_0_LS}
    \Do
        \If{$ n = 1$}
             \State Derive $\boldsymbol{\theta}^0$ from $\boldsymbol{\boldsymbol{p}^0} $ (see \eqref{eq:a_min_a_max} -- \eqref{eq:R_min_R_max})
        \Else 
            \State Derive $\boldsymbol{\theta}^0$ from ${\hat{\boldsymbol{p}}^{n-1}} $
        \EndIf
            \State Solve \eqref{eq:optimization_prob_reform} for $\mathbf{\tilde{v}}, \mathbf{\tilde{i}}$ and save the estimate $\boldsymbol{\hat{\theta}}$
            \State Reconstruct the LF parameters $\boldsymbol{\hat{p}}$ from $\boldsymbol{\hat{\theta}}$
        \If{$n > 1$}
            \State $\Delta = \left|\frac{\boldsymbol{\hat{\theta}}^n-\hat{\boldsymbol{\theta}}^{n-1}}{\hat{\boldsymbol{\theta}}^{n}}\right|$
        \EndIf
             \State $n\gets n + 1$
    \doWhile{($\Delta>\varepsilon$ \textbf{or} $n \leqslant n_{RC}^{\max}$)}
    \State $\boldsymbol{\hat{\theta}}^\star =\boldsymbol{\hat{\theta}}^{n-1}$  $\boldsymbol{\hat{p}}^\star =\boldsymbol{\hat{p}}^{n-1}$, 
    \State \Return $\boldsymbol{\hat{p}}$
\EndProcedure
\end{algorithmic}
\end{algorithm}

\clearpage
\newpage
\section{Results}\label{sec:results}

This section presents the results of a numerical study conducted to validate the proposed method for estimating low-frequency (LF) parameters from time-domain measurements. Unlike conventional approaches in the literature, which typically rely on predefined excitation signals such as pulses, multisine inputs, or pseudo-random binary sequences (PRBS), the present study considers a more application-oriented scenario. Specifically, the battery is subjected to an excitation profile representative of frequency containment reserve (FCR) in power systems.

The motivation associated with this particular application of frequency containment reserve (FCR) stems from the fact that the provision of FCR is continuous, namely, the battery never rests, making the ECM parameters assessment technically difficult. As a matter of fact, battery providing grid ancillary services usually require parameters inference techniques that use the operational states of the asset to feed a suitable ECM estimation technique.

\subsection{Frequency containment reserve power profile}
The power profile applied to the battery corresponds to a typical FCR. The profile is obtained using historical data of the measured grid frequency of continental Europe, $f(k)$, $k = 1,...,T$, sampled at $T_s = 1~\text{s}$. The power profile is obtained as follows \cite{noauthor_frequency_nodate}, \cite{noauthor_commission_2017}:
\begin{equation}
    P_{FCR}(k) = \begin{cases}
    -P_{FCR}^{\max},            &\Delta f(k)\leqslant -0.2~\text{Hz}\\
        \varsigma\cdot\Delta f(k), & |\Delta f(k)|<0.2~\text{Hz}\\
    P_{FCR}^{\max},             & \Delta f(k) \geqslant 0.2~\text{Hz}\\
    \end{cases}
\end{equation}
where $P_{FCR}^{\max}$ is the maximum power in case of full activation of the FCR, if the frequency deviation $\Delta f$ exceeds 0.2~Hz, and $\varsigma$ is the droop in W/Hz.

\subsection{Synthetic BMS Data Generation}

\begin{algorithm}[htb!]
\caption{Generation}
\begin{algorithmic}[1]
\Procedure{Generate Data}{$\mathbf{i}$}
    \State Initialize $f_{\max}^0, \Delta_\phi, \boldsymbol{p}, T_s , n, \text{SOC}_0, \alpha, \beta, C_\text{nom}$
    \State Decompose $\boldsymbol{p}$ to obtain $\boldsymbol{\theta}$ \eqref{eq:a_and_b_definition}--\eqref{eq:gain_definition}
    \State Obtain $\mathbf{v}$ from \eqref{eq:ss_discrete}
        \State $\Delta \mathbf{v}\gets \mathcal{N}(0,\sigma_V^2)$
        \State $\Delta \mathbf{i}\gets \mathcal{N}(0,\sigma_I^2)$
        \State $\tilde{{\mathbf{v}}} = \mathbf{v} + \Delta \mathbf{v} $
        \State $\tilde{{\mathbf{i}}} = \mathbf{i} + \Delta \mathbf{i} $
    \State \Return $\mathbf{\tilde{v}}, \mathbf{\tilde{i}}$
\EndProcedure
\end{algorithmic}
\label{alg:Data_generation}
\end{algorithm}

In the numerical study where we have access to the true parameter values of non-approximated ECM, $\boldsymbol{p}$, we first artificially generate the output data, $v $ for a given the input power excitation of the battery under study. Having the battery power, we first translate it to voltage and current profiles by using the parameters of the ECM and assuming the initial voltage to be equal to the OCV after battery was at rest and iteratively deviding the power with voltage to obtain the current. 
The current excitation corresponds the FCR profile on 1 Feb 2026 with total duration $T = 3 $~h and the sampling time $T_s = 1$~s, hence $N = 10800$ points. The approximated ECM is obtained by applying \eqref{eq:ss_discrete} using the true parameters $\boldsymbol{\theta}$, obtained by decomposition of CPE after applying \eqref{eq:a_and_b_definition}--\eqref{eq:gain_definition} for a large number of branches ($n = 100$). The maximum frequency of interest used for the decomposition is $f_{\max} = f_s/\pi\approx 0.3183$~Hz, which is derived in accordance to the forward-Euler discretisation stability limit (see the Appendix \ref{app:Appendix_Stability_limit}) and the maximum phase ripple is set to $\Delta_\phi = 0 $. Parameters are assumed to be constant for SOC intervals  $\left[(k-1)\cdot 0.1, \; k\cdot0.1\right]$  (unitless), where $k=1,...,10$. The initial SOC is $\text{SOC}_0=0.55$. The FCR droop is set to $\varsigma = 100 $~W/Hz, for which the SOC remains within the range between 0.5 and 0.6 where the linearized OCV segment, given by \eqref{eq:E_SOC_linear}, has coefficients $\alpha_6 = 3.18 $~V and $\beta_6 = 1 $~V.

\begin{table}[htb!]
\centering
\caption{True parameter values of the approximated ECM associated to an NMC 60 Ah Li-ion cell.}
\label{tab:true_parameters}
\begin{tabular}{cccc}
\toprule
$R_\Sigma\,(\Omega)$    & $ Q\,(\text{S}\cdot \text{s}^{\phi})$     & $\phi \,(-)$ & $f_{\max}\,(\text{Hz})$ \\
\midrule
0.0014   &  22281  & 0.52 & $\frac{f_s}{\pi}\approx0.3183 $\\
\bottomrule
\end{tabular}
\end{table}

\noindent The input and output variables (i.e., cell's current and voltage) are then corrupted with independent Gaussian noise with zero mean and standard deviation $\sigma_V = \frac{1}{3}\cdot 0.0015 $ V and $\sigma_I = \frac{1}{3}\cdot0.05 $ A, which are the typical values for a conventional BMS accuracy. This step is needed in order to represent measurement noise of these two cell's variables. The data generation process is summarised in pseudo-code in Algorithm~\ref{alg:Data_generation}.

To test the impact of the solution space reduction, we impose the upper and lower bounds on CPE parameters, $\phi$ and $Q$ from $\boldsymbol{p}$, as well as the range of maximum frequency $f_{\max}$.

\subsection{Sensitivity to Initial Guess and Parameter Bounds}

We perform simulations in which the fixed excitation current and the generated output voltage (using the true ECM parameters) are corrupted $N_{\text{sim}}=1000$ times with the noise of the same statistical properties. The collection of noisy data is then used to estimate the parameters.  The decomposed ECM parameters are initialised from uniformly distributed randomly chosen CPE parameters, $\phi$ and $Q$, within the corresponding lower and upper imposed limits, $[\phi_{\min}, \phi_{\max}]$ and $[Q_{\min}, Q_{\max}]$, respectively. The value for the maximum frequency is chosen as a mid-point of the interval $\left[\frac{f_s}{N}, \frac{f_s}{\pi}\right)$:
\begin{equation}
    f_{\max}^0 = \dfrac{1}{2}\left(f_{\max}^\downarrow(n) + f_{\max}^\uparrow(n)\right) = \dfrac{f_s}{2}\left(\frac{1}{\pi} + \frac{1}{N}\right)
\end{equation}
We consider three cases, for which the limits for CPE parameters are listed in Tab.~\ref{tab:cases_lb_ub}. Case 1 allows a wide range for parameters $\phi$ and $Q$ as well as for the frequency $f_{\max}$, whereas in Case 2 and 3 all the intervals are gradually shrunk, reducing the solution space.

\begin{table}[htb!]
\centering
\caption{Three considered cases for parameter upper and lower bounds.}
\label{tab:cases_lb_ub}
\begin{tabular}{lccc}
                         & Case 1         & Case 2   & Case 3      \\
\midrule
{[}$\phi_{\min}$, $\phi_{\max}${]} & {[}0.30, 0.70{]} & {[}0.40, 0.60{]} & {[}0.45, 0.55{]} \\
{[}$Q_{\min}$, $Q_{\max}${]~$(\text{S}\cdot \text{s}^{\phi})$  }       & {[}1e2, 1e6{]} & {[}1e3, 1e5{]} & {[}1e4, 3e4{]}\\
\bottomrule
\end{tabular}
\end{table}

\noindent The statistical results are summarised in Tab.~\ref{tab:parameter_errors_combined} and show that by executing the Algorithm 1, the estimates have almost unchanged accuracy. This means that the estimation process is robust to the initial guess and even wide parameter bounds do not affect the estimator accuracy.
On average, the accuracy of all retrieved parameters, including CPE, are bellow 1~\%, namely $0.19~\%$ for $R_{\Sigma}$, $0.77~\%$ for $Q$ and $0.17~\%$ for $\phi$. Minimum error shows excellence performance in certain instances where the estimate parameter values are almost equal to the true ones. Maximum error is highest for $Q$ in Case 1, with still relatively low value or $3.6~\%$, whereas errors of $R_{\Sigma}$ and $\phi$ are bounded by $0.78~\%$. 

\begin{figure}[htb!]
    \centering
    \includegraphics[width=1.0\linewidth]{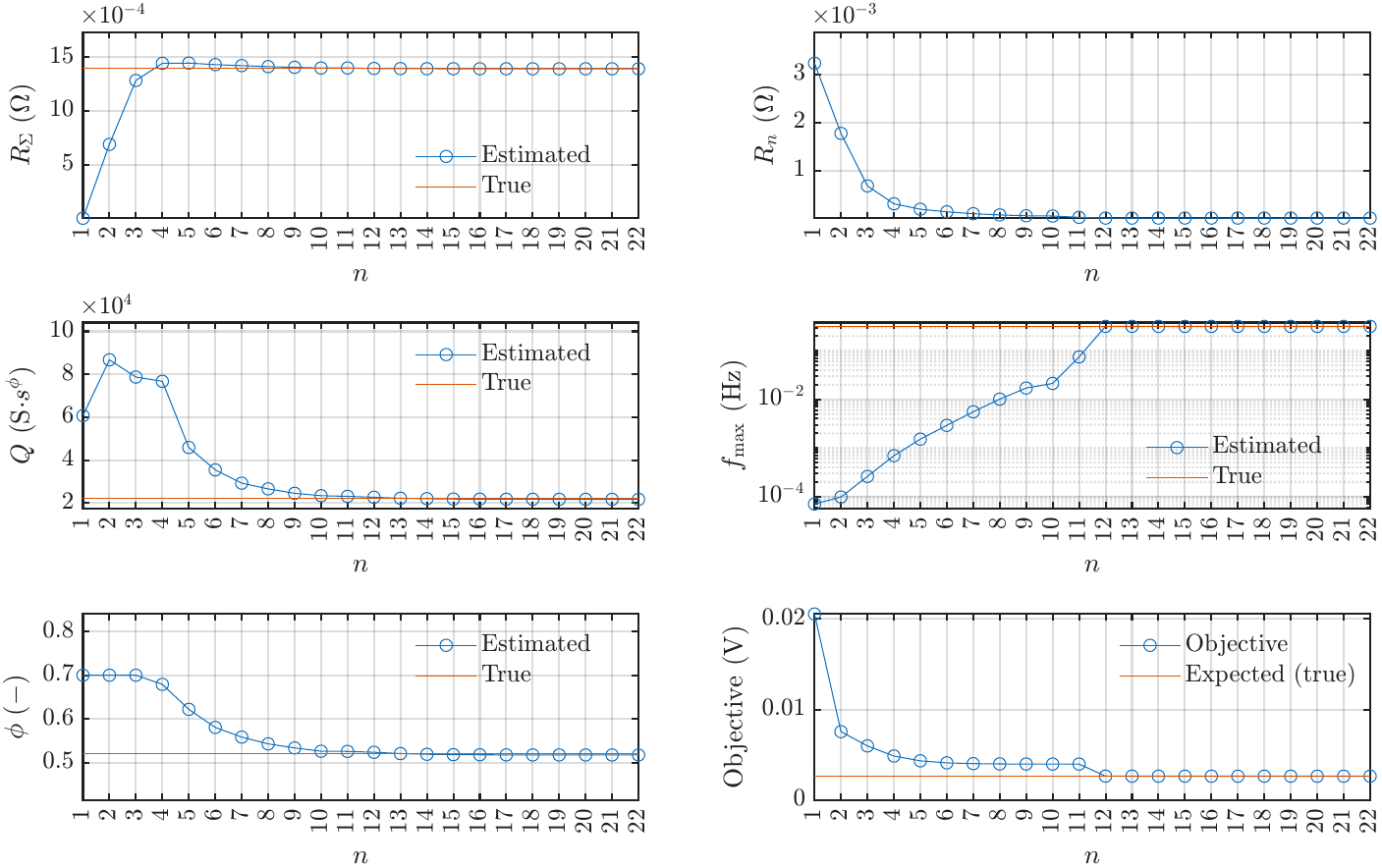}
    \caption{Evolution of estimated parameters $\boldsymbol{p}$ along iterations of Alg.~1 (left) and $R_n$, $f_{\max}$ and the objective function value (right). The stopping criteria are satisfied for $n=22$ RC branches.}
    \label{fig:Iterations}
\end{figure}

The evolution of estimated parameters, $\boldsymbol{\hat{p}}$, along the iterations of the Alg.~\ref{alg:Estimation_proc} is shown in Fig.~\ref{fig:Iterations}. In addition, we show the evolution of $R_n$ (corresponding to the RC branch with the lowest time constant), $f_{\max}$ and the objective function value. Given the values for $\varepsilon = 1\times10^{-4}$ and $n_{RC}^{\max} = 25$, the algorithm converges after 22 iterations, satisfying the stopping criteria.

\begin{table*}[h!]
\centering
\caption{Comparison of true parameters ($\boldsymbol{p}$), estimated parameter mean values ($\mu_{\boldsymbol{\hat{p}^\star}}$), standard deviations ($\sigma_{\boldsymbol{\hat{p}^\star}}$), and associated error metrics for Cases 1--3.}
\begin{subtable}[t]{1\linewidth}
\centering
\begin{tabular}{lcccccc}
\hline
 & $\boldsymbol{p}$ &  $\mu_{\boldsymbol{\hat{p}^\star}}$ &  $\sigma_{\boldsymbol{\hat{p}^\star}}$ &  $\mu_{\delta}~(\%)$ & $\delta_{\min}~(\%)$ &  $\delta_{\max}~(\%)$\\ 
\hline
$R_\Sigma~(\Omega)$  & 0.0014 & 0.0014 & 3.16E-06 & 0.18537 & 0.00016 & 0.71528 \\
$Q~(\text{S}\cdot \text{s}^{\phi})$    & 22281.00  & 22237.45  & 206.4300 & 0.74403 & 0.00034 & 3.64595 \\
$\phi~(-)$ & 0.5200 & 0.5197 & 0.0010 & 0.15928 & 0.00046 & 0.80464 \\
\hline
\end{tabular}
\caption{Case 1: Wide parameter bounds.}
\label{tab:params_case1}
\end{subtable}

\vspace{0.75em}

\begin{subtable}[t]{1\linewidth}
\centering
\begin{tabular}{lcccccc}
\hline
 & $\boldsymbol{p}$ &  $\mu_{\boldsymbol{\hat{p}^\star}}$ &  $\sigma_{\boldsymbol{\hat{p}^\star}}$ &  $\mu_{\delta}~(\%)$ & $\delta_{\min}~(\%)$ &  $\delta_{\max}~(\%)$\\ 
\hline
$R_\Sigma~(\Omega)$  & 0.0014 & 0.0014 & 3.19E-06 & 0.18315 & 0.00102 & 0.77900 \\
$Q~(\text{S}\cdot \text{s}^{\phi})$ & 22281.00 & 22240.83 & 208.2533 & 0.77076 & 0.00021 & 2.91933 \\
$\phi~(-)$ & 0.5200 & 0.5197 & 0.0011 & 0.16836 & 0.00045 & 0.64832 \\
\hline
\end{tabular}
\caption{Case 2: Medium parameter bounds.}
\label{tab:params_case1}
\end{subtable}

\vspace{0.75em}

\begin{subtable}[htb!]{1\linewidth}
\centering
\begin{tabular}{lcccccc}
\hline
 & $\boldsymbol{p}$ &  $\mu_{\boldsymbol{\hat{p}^\star}}$ &  $\sigma_{\boldsymbol{\hat{p}^\star}}$ &  $\mu_{\delta}~(\%)$ & $\delta_{\min}~(\%)$ &  $\delta_{\max}~(\%)$\\ 
\hline
$R_\Sigma~(\Omega)$  & 0.0014 & 0.0014 & 3.27E-06 & 0.18876 & 0.00038 & 0.74843 \\
$Q~(\text{S}\cdot \text{s}^{\phi})$ & 22281.00 & 22253.28 & 214.5470 & 0.77432 & 0.00072 & 3.41157 \\
$\phi~(-)$ & 0.5200 & 0.5198 & 0.0011 & 0.16622 & 0.00029 & 0.70518 \\
\hline
\end{tabular}
\caption{Case 3: Narrow parameter bounds.}
\label{tab:params_case1}
\end{subtable}

\label{tab:parameter_errors_combined}
\end{table*}
\noindent 

\subsection{Hypothesis verification of the measurement model}
\begin{figure*}[htb!]
    \centering
    \begin{subfigure}[t]{0.53\textwidth}
        \centering
        \includegraphics[width=\linewidth]{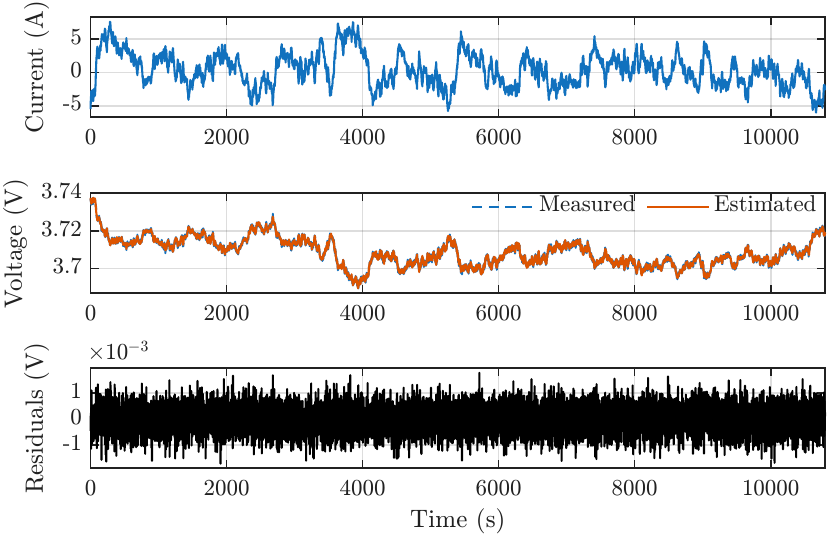}
        \caption{}
        \label{fig:sub_timeseries}
    \end{subfigure}
    \hfill
    \begin{subfigure}[t]{0.4\textwidth}
        \centering
        \includegraphics[width=\linewidth]{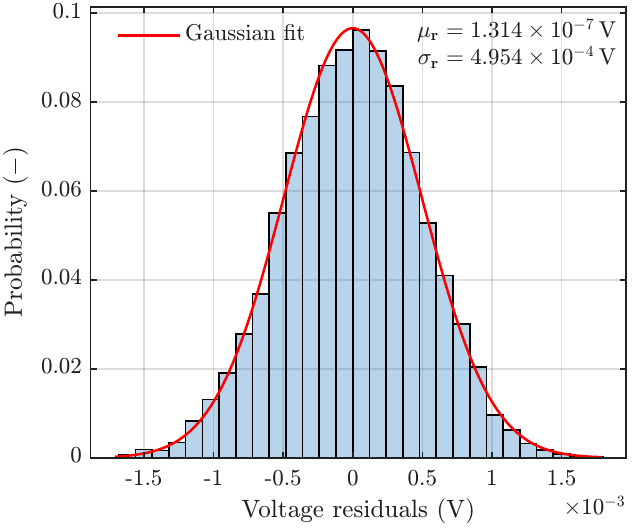}
        \caption{}
        \label{fig:sub_hist}
    \end{subfigure}
    \caption{(a) Current, voltage (measured and reconstructed) profiles and voltage residuals (b) and voltage residual distribution.}
    \label{fig:Measured_vs_reconstructed}
\end{figure*}

After estimating the parameters, we can reconstruct the voltage response by executing the state-space model. Fig.~\ref{fig:sub_timeseries} shows the reconstructed and measured voltage. The computed residuals, namely:
\begin{equation}
    \mathbf{r} = \tilde{\boldsymbol{y}}- {\boldsymbol{y}}(\boldsymbol{\hat{\theta}}^\star)
\end{equation}
are shown as well.
From the plot we can conclude that residuals, for the great majority of the points, do not exceed $1.5$ mV, by absolute value. The histogram of the residuals is shown in Fig.~\ref{fig:sub_hist}. The residuals follow normal distribution, and the computed mean and standard deviations are:
\begin{equation}
    \mu_{\mathbf{r}} = 1.314\times 10^{-7}~\text{V} \quad \text{and} \quad \sigma_{\mathbf{r}} = 0.495 \times 10^{-3}~\text{V}\approx \sigma_V=0.500 \times 10^{-3}~\text{V}
\end{equation}
This certifies that the estimator is unbiased and that the standard deviation of voltage residuals matches well the instrument accuracy defined by $\sigma_V$.

\clearpage
\newpage
\section{Conclusion}\label{sec:Conclusion}

In this paper, we addressed the problem of identifying observable low-frequency (LF) parameters of equivalent circuit models (ECMs) using time-domain measurements obtained from low-frequency sampled data provided by built-in battery management systems (BMS) during operation. Unlike conventional approaches that rely on artificially imposed excitation signals, requiring the battery to be offline, and low-order models, the proposed method leverages naturally occurring operational data and a higher-order RC network with a recursive definition to approximate the constant phase element (CPE) behavior in time domain.

The identification problem is formulated as a constrained, non-convex least-squares optimization with constraints written as a discretized state-space model. To improve tractability, we introduced parameter initialization and imposed bounds, reducing the effective search space. Additionally, an algorithmic procedure is proposed to determine a sufficient number of RC branches to improve the convergence.

The results of the numerical study where we considered a power system application where battery is providing frequency containment reserve, we used corresponding voltage and current data while battery, shows that the proposed approach, under noise levels typical for conventional BMS, ensures accurate LF parameters estimates with accuracy less than $1~\%$ for each parameter, on average. In worst case, from 1000 simulations, the maximum error of estimate is $3.6~\%$ for $Q$ while for $R_{\Sigma}$ and $\phi$ the maximum errors remain bellow $1~\%$, namely $0.78~\%$ and $0.80~\%$, respectively.

Thanks to the continuous tracking of the evolution of these parameters over time, provided by the proposed method, the modeller gets the access to improved understanding of battery degradation processes. Relying only on measurements available from built-in BMS, the approach avoids intrusive testing and specialized laboratory equipment, making it suitable for online implementation and can be used in scope of ageing-aware control strategies.

\clearpage
\newpage

\appendix
\section{Stability limit of maximum frequency for the decomposition}
\label{app:Appendix_Stability_limit}
\numberwithin{equation}{section}
\setcounter{equation}{0}

In a discrete-time state-space implementation of an RC branch discretisation, stability imposes an upper limit on how fast the branch dynamics (i.e., how large its characteristic frequency) is relative to the sampling frequency, $f_s = \dfrac{1}{T_s}$.
In \eqref{eq:optimization_prob}, the equations are written as a standard, discretised, state-space model, using the forward Euler discretisation method. To ensure the stable response, all the eigenvalues of the matrix $\mathbf{A}_d = \mathbf{I} + T_s\cdot \mathbf{A}_c$ must be within a unit circle:
\begin{equation}\label{eq:eigen_within_unit}
    |\lambda_i(\mathbf{A}_d)|<1,\quad\forall i = 1,...,n
\end{equation}
Due to its diagonal structure, the eigenvalues are obviously
\begin{equation}\label{eq:eigen_formulas}
    \lambda_i(\mathbf{A}_d)=1-T_s\cdot\dfrac{1}{R_i C_i}
\end{equation}
and therefore, according to \eqref{eq:eigen_within_unit}:
\begin{equation}
    \left|1-T_s\cdot\dfrac{1}{R_i C_i}\right|<1,\quad\forall i = 1,...,n
\end{equation}
\begin{equation}
    -1<1-T_s\cdot\dfrac{1}{R_i C_i}<1,\quad\forall i = 1,...,n
\end{equation}
which is transformed into two inequalities:
\begin{equation}
    T_s\cdot\dfrac{1}{R_i C_i}<2\quad \wedge\quad T_s\cdot\dfrac{1}{R_iC_i} >0
\end{equation}
where the second one is always satisfied, and the first one can be written in terms of characteristic frequencies of RC branches:
\begin{equation}
    2\pi f_{c_i} < 2f_s \Longleftrightarrow f_{c_i}<\dfrac{f_s}{\pi}\quad  \forall i = 1,...,n
\end{equation}
which is equivalent to
\begin{equation}
    \max_i f_{c_i} = f_{\max} < \dfrac{f_s}{\pi}.
\end{equation}

\newpage

\section*{Acknowledgements}
This research is carried out in the frame of Swiss Circular Economy Model for Automotive Lithium Batteries (CircuBAT) Flagship project, with the financial support of the Swiss Innovation Agency (Innosuisse - Flagship Initiative) (FLAGSHIP PFFS-21-20).
\pagebreak

\clearpage
\bibliographystyle{IEEEtran}
\bibliography{references}
\end{document}